\documentclass[tighten, times, twocolumn]{aastex631}  % ApJ look-alike
\shorttitle{Radio emission cross the entire rotation phases of pulsars}
\shortauthors{Wang \& Han}

\begin{document}
%\linenumbers
%\title{Curvature radiation model for the whole phase emission of radio pulsars}
\title{Radio emission across the entire rotation phases of pulsars}

%% AUTHOR/INSTITUTIONS FOR AASTEX6.1:
\author{P.~F.~Wang}
\affiliation{National Astronomical Observatories, Chinese Academy of Sciences, 20A Datun Road,
     Chaoyang District, Beijing 100101, China}
\affiliation{School of Astronomy, University of Chinese Academy of Sciences,
     Beijing 100049, China}

\author{J.~L.~Han}
\affiliation{National Astronomical Observatories, Chinese Academy of Sciences, 20A Datun Road,
     Chaoyang District, Beijing 100101, China}
\affiliation{School of Astronomy, University of Chinese Academy of Sciences,
     Beijing 100049, China}

\email{pfwang@nao.cas.cn; hjl@bao.ac.cn}

% %% Extra info for aastex:
% \received{Yesterday}
% \revised{Today}
% \accepted{Tonight}
% \published{Tomorrow}
% \submitjournal{AASJournal}

\begin{abstract}
Super-sensitive observations of bright pulsars by the Five-hundred-meter Aperture Spherical radio Telescope (FAST) have revealed weak radio emission continuously emerged in the rotation phases between the main pulse and interpulse of an rotating neutron star. We develop a model for the polarized radio emission radiated from different heights in the pulsar magnetosphere and examine emission intensity distribution over the whole rotation phases of pulsars seen from all directions by the line of sight. We find that for pulsars with small periods and the magnetosphere filled with much more relativistic particles, the polarized radio emission can be generated in all rotation phases for both the aligned and perpendicular rotating neutron stars. When the line of sight cuts the pulsar emission beam between the rotation and magnetic axes, the polarization angles have the same sense of variation gradient for the ``main'' pulse and ``interpulse''. If the line of sight cuts the beams between the inclined magnetic axis and the equator, the opposite senses can be found for the main pulse and interpulse. In addition to the pulsed emission, we find persistent radio emission generated in the pulsar magnetosphere. The model can naturally explain the emission across the entire rotation phases. 
\end{abstract}

% http://journals.aas.org/authors/keywords2013.html
\keywords{pulsars: general}

%==================================
\section{Introduction}
\label{introduction}

%%%%%%%%%%%%%%%%%%%%%  observations
The mean polarization profiles of pulsars reflect the emission geometry and emission processes in the magnetosphere of neutron stars. Radio observations can give details of polarization profiles \citep[e.g.][]{gl98, hdvl09, pkj+23, whx+23}. In general, pulsars have profiles with a duty cycle of about several percent of the rotational phase. Some pulsars have wide profiles with emission in the phase of more than 180$^\circ$ and even full of all the rotation phases up to almost 360$^\circ$, such as PSRs J1916+0748 and J2007+2722 \citep{whx+23}. The profiles typically have one, two, three, or more components, and are often linearly and circularly polarized. The polarization position angles of some pulsars exhibit a S-shaped curve or have orthogonal modes at some phases. In addition to conventionally detected pulses, some pulsars may have persistent radio emission as revealed by the aperture synthesis mapping observations \citep{grf+22}. 

%%%%%%%%%%%%%%%%%%%%% empirical models
Some empirical or physical models have been proposed to understand diverse pulsar profiles. The observed polarization profiles are related to pulsar geometry \citep{rc69}, emission mechanisms \citep[e.g.][]{xlhq00}, propagation effects \citep[e.g.][]{ab86,wlh10,bp12} or their combination \citep{wwh14}. They result from the cutting of pulsar emission beams by sight lines. The beams may have a shape in circular \citep{lm88,gh96}, or elongated in the meridional plane direction \citep{nv83}, or compressed along which \citep{big90, qlw+04, gan04}. Profile components originate from the radiation beams that might have configurations of a conal and/or core \citep{ran83}, or patchy \citep[e.g.][]{lm88,hm01}, or fan-shaped \citep[e.g.][]{wpz+14}. It is very hard to reveal the real shape and configuration of the emission beam of one pulsar. The procession of the rotation axis of some binary pulsars, such as PSR J1906+0746, indeed provides us a chance to figure out some part of the emission beam \citep{dkl+19} by using observations over many years.

Curvature radiation is one of the most probable mechanisms for pulsar radiation, and has been taken to interpret pulsar polarization\citep[e.g.][]{gs90}. 
%%%%%%%%%%%%%%%%%%%%%  theoretical models and model predictions
When the detailed emission geometry and particle distributions are taken into account, the swing of polarization position angle curves and circular polarization can be naturally produced, as shown by simulations \citep{gan10, wwh12}. Previously, most pulsar emission models are confined to emission generated from a fixed height of pulsar magnetosphere and detected by a given sight line. The polarized emission from the entire magnetosphere has not yet been explored.

%++++++++++++++++++++++++++
\begin{figure*}
  \centering
  \includegraphics[width = 0.48\textwidth] {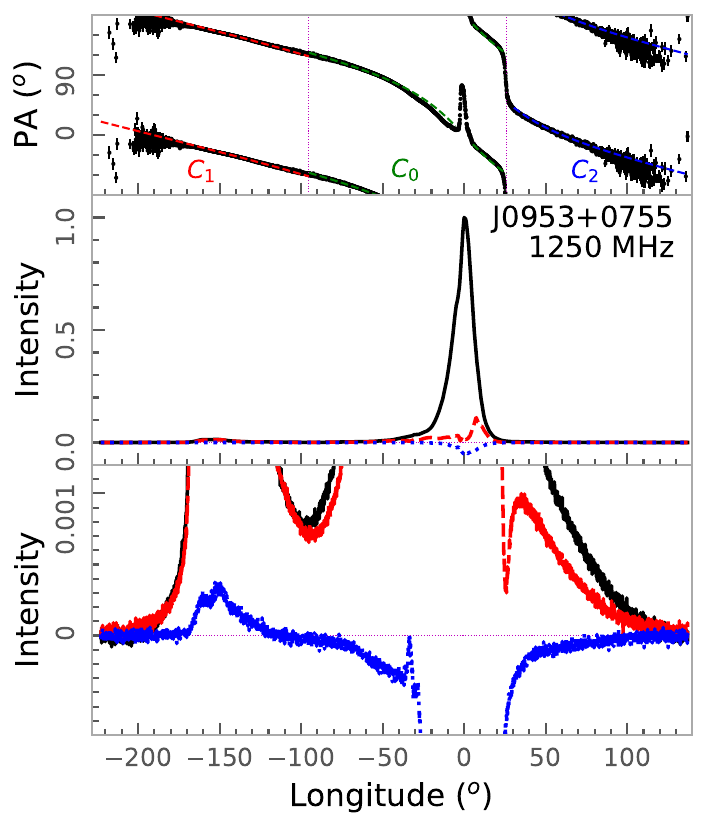}
  \includegraphics[width = 0.48\textwidth] {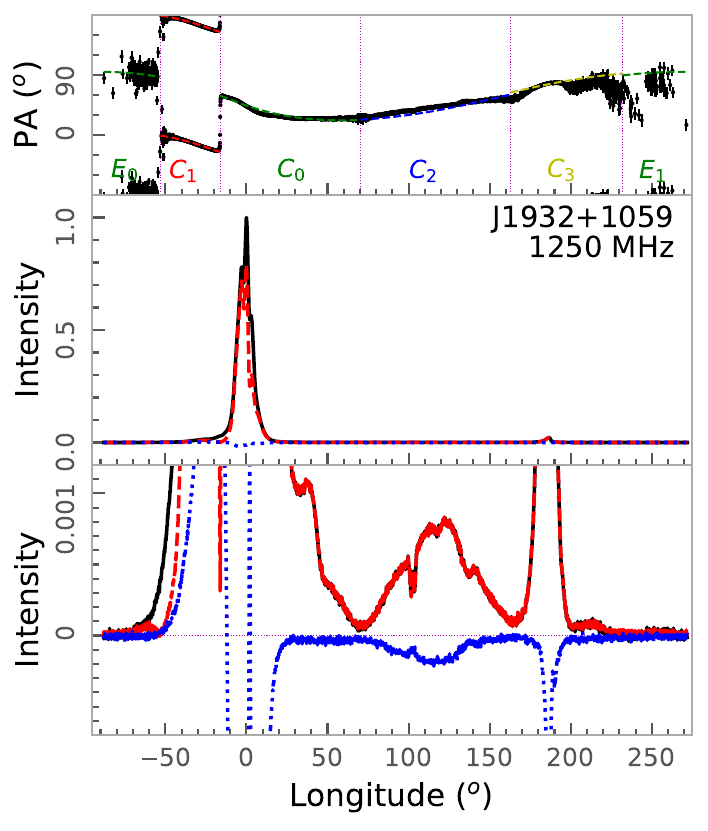}\\
  \caption{Mean polarization profiles of PSRs J0953+0755 and J1932+1059. The total intensity, linear and circular polarization intensities are represented by solid line, dashed line and dotted line in the bottom and middle sub-panels for each pulsar, while the bottom sub-panels are zoomed to show weak parts of profiles. Polarization position angles are shown in the top sub-panels by dots with error-bars for data of linear polarization intensity greater than 3$\sigma$. The piecewise RVM fitting is performed to the PAs in three phase ranges ($C_0$, $C_1$ and $C_2$, as separated by the vertical dashed lines) for PSR J0953+0755 with orthogonal mode jump modelled for $C_2$, and four ranges ($C_0$, $C_1$, $C_2$ and $C_3$) for PSR J1932+1059 with $C_1$ having orthogonal mode jump modelled. The two ranges ($E_0$ and $E_1$) with very weak emission and PAs with large uncertainties for PSR J1932+1059 are not modeled, but the predicted PA curve is plotted. }
  \label{fig:Obs_profs}
\end{figure*}
%------------------------------
In this paper, we develop a model by accounting for the curvature radiation from the entire pulsar magnetosphere to explain the weak emission detected from the whole phase of pulsar rotation. We first present the sensitive FAST observations of bright pulsars to show the weak emission generated within one pole or between two poles of the beam in Section 2. In Section 3, we set the model assumptions, and present the simulation methods and results from a static pulsar magnetosphere. In Section 4, we compare the model to observations. Discussion and conclusions are given in Section 5.

\begin{figure}
  \centering
  \includegraphics[width = 0.45\textwidth] {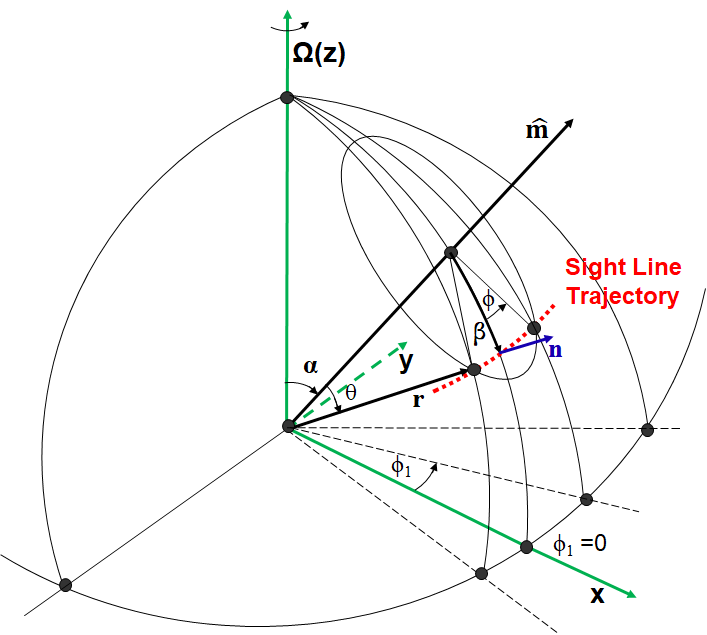}
  \caption{Geometry of pulsar magnetosphere. The magnetic axis {\bf m} is
    inclined at an angle of $\alpha$ with respect to the rotation axis
    $\Omega$. The line of sight $\mathbf{n}$ cuts across the emission beam with an impact angle $\beta$ from the magnetic axis, whose trajectory is indicated by the dotted arc. The line of sight always has an angle of $\zeta=\alpha+\beta$ with respect to the rotation axis, so that it is named as being the sight line angle. We use $\alpha_0=180^\circ-\alpha$, $\beta_0=-\beta$ and $\zeta_0=180^\circ-\zeta$ to mark the values derived from observations. A given emission position in the magnetosphere can be described by three parameters: $\mathbf{r}$, $\theta$ and $\phi$ with respect to the magnetic axis frame. In the rotation axis frame with the z-axis along the rotation axis, the x-axis lies in the meridional plane defined by the magnetic and rotation axes and the rotation phase is indicated by $\phi_1$.}
  \label{fig:geo}
\end{figure}

%-------------------------------------
\begin{table}
  \centering
  \caption{Geometry parameters of PSRs J0953+0755 and J1932+1059.}
  \label{table:geo}
%  \tabcolsep 3pt
  %\scriptsize
   \begin{tabular}{cchcc}
    \hline
    \hline
%\hline
%PSR        & Period  & $W$$^\dagger$   & 
$\alpha_0$ & $\beta_0$ & $\zeta_0$ & Freq.  & Ref.\\
%           &  (s)    &($^\circ$) &
           ($^\circ$)  & ($^\circ$) & ($^\circ$) & (GHz)  &     \\
\hline
\multicolumn{5}{c}{PSR J0953+0755} \\ %& 0.253   & 316.2(1)& 
\hline
170       &  5        &  - & 0.43 & 1 \\
174.1     &  4.2      &  - & 0.4  & 2 \\
174(90)   &  2.5(40)  &  - & 0.43 & 3 \\
174(30)   &  2.5(15)  &  - & 1.42 & 3 \\
168       &  8.5      &  - & 1.0  & 4 \\
179.3(150)&  1.0(150) &  - & 1.41 & 5 \\
153.1(900)&  4.1(50)  &  - & 4.85 & 5 \\
105.4(5)  & 22.1(1)   &  - & 1.42 & 7 \\
\hline
\multicolumn{5}{c}{PSR J1932+1059} \\ %& 0.226   & 272.9(1) & 
\hline
35      & 23     & - & 0.43  & 1 \\
6       & 4      & - & 0.4   & 2 \\
25(2)   & 16(2)  & - & 0.43  & 3 \\
27(4)   & 16(3)  & - & 1.42  & 3 \\
90      & 41.8   & - & 1.0   & 4 \\
41(8)   & 21(4)  & - & 0.41  & 6 \\
51(3)   & 35(3)  & - & 0.61  & 6 \\
61(2)   & 39(2)  & - & 1.41  & 6 \\
36(1)   & 25.6(9)& - & 1.42  & 7 \\
\hline
\end{tabular}
%\tablecomments{$^\dagger$profile width at 1/10000 the peak intensity level. $\alpha_0$ and $\beta_0$ follows the convention of \citet{ew01}.
%$^\ddagger$ geometry is referred to a longitude of $354^\circ$.} 
\tablerefs{1: \citet{nv82}; 2: \citet{lm88}; 3: \citet{bcw91}; 4: \citet{ran93}; 5: \citet{vx97}; 6: \citet{stc99}; 7: \citet{ew01}; 
%8: \citet{akc+13}
}
\end{table}

%==================================
\section{Radio emission over wide rotation phases of two pulsars}
\label{sec:observations}

Here we consider two bight pulsars, PSRs J0953+0755 and J1932+1059, which have been observed by the Five-hundred-meter Aperture Spherical radio Telescope (FAST) \citep{nan06}. The weak polarized emission in a wide rotation phase range or the entire phase between the two magnetic poles has been well-detected. 

The bright PSR J0953+0755 is one of the first few pulsars ever discovered \citep{phb+68}, which has a spin period of 0.253s and has the emission detected in almost all the rotation phases in the frequencies from 55 MHz to 1425 MHz \citep[e.g.][]{hc81,mr04,bgp+22,wlj+22}. There was no consensus on the emission coming from the beam of one magnetic pole \citep{nv83,lm88,bcw91,ran93} or two poles \citep{ew01}. 

PSR J1932+1059 is also very bright. It was discovered during Molonglo pulsar search\citep{lvw68} and has a spin period of 0.226~s. The main pulse and the interpulse are separated by 180$^{\circ}$ which is most likely generated from two poles \citep{rr97}. The bridge emission between the main pulse and inter-pulse has been detected at 408, 430, 1170, 1250 and 1665MHz \citep{pl85,phi90,mr04,kyp+21}. Because of the bridge emission, it has been argued to originate from a single pole \citep[e.g.][]{lm88,phi90}.

Emission geometries of these two pulsars have been investigated through either the empirical beam-analysing method \citep{lm88, ran93} or fitting to the polarization position angle curves with the rotating vector model (RVM) \citep[e.g.][]{phi90,bcw91,ew01}. The geometric parameters derived by previous authors (and also this work, see below) are listed in Table~\ref{table:geo}.

The FAST observations of these two pulsars were carried out on 2022 August 29th and 2022 November 21th respectively for 6442 seconds each by using the central beam of the L-band 19-beam receiver. The receiver covers the frequency range of 1.0-1.5~GHz, and has a system temperature of about 22~K \citep{jth+20}. Signals from the receiver were digitally channelized to 4096 frequency channels, and then sampled at a time resolution of 49.152$\mu s$ for the 4 polarization channels, $XX$, $YY$, Re[$X^{*}Y$] and Im[$X^{*}Y$]. The recorded FAST data were dedispersed and folded.  Polarization calibrated following the procedures described by \citep{whx+23}. In short, the gain, differential gain and differential phase in the two polarization $X$ and $Y$ are calibrated by using injected noise diode signals recorded before or after the observation session. Because the receiver has the XY feed, the polarization leakage has been tested and found to be less than 0.03\% \citep{clh+22}. 

We obtained the mean polarization profiles of the two pulsars as shown in Figure~\ref{fig:Obs_profs}. They come from the most sensitive observations ever available and have a dynamical range of more than 20000.  The profile details at a level of 1/10000 the peak intensity can even be detected with a very high signal-to-noise ratio (S/N), which locate at the rotation phases between the main pulse and inter-pulse\footnote{Here, we set the lowest points of the profiles as the zero-level for the baseline, assuming that the flux densities there go to zero. It is possible that there is still emission at these phases and even polarized. If so such a possible offset would distort the presented position angles more severely for the lower level polarized emission.}. Similar results have also been shown by \citet{wlj+22} and \citet{kyp+21}.

PSR J0953+0755 has a wide profile over 310$^\circ$ phase at the level of 1/10000 the peak intensity, as shown in Figure~\ref{fig:Obs_profs}. The interpulse is about 150$^\circ$ ahead of the main pulse, with an intensity of 2\% the main pulse peak. The continuous bridge emission between the main pulse and the interpulse is highly polarized, and has an intensity of only about 1/1000 the main pulse, but clearly detected by FAST. Compared with previous results obtained from Arecibo observations \citep{mr04}, the new profiles in Figure~\ref{fig:Obs_profs} show the weak interpulse emission at the longitude of $-150^\circ$ and even weaker bridge emission between the two peaks, extending almost 360$^\circ$ of the rotation phases.  %The geometry parameters estimated by previous authors are listed in  Table~\ref{table:geo}.  
Because emission at different phases originates from various heights in the pulsar magnetosphere, the rotation-induced distortion of polarization position angles therefore can be different \citep[e.g.][]{dy08,wwh12} and the traditional RVM is hard to  describe the PA variations over the whole rotation phase. We therefore develop a new piecewise RVM, as described in the Appendix~\ref{sec:method}, to fit the PA curve. A shown in Figure~\ref{fig:Obs_profs}, the piecewise RVM can fit the observed PA values well. The precise geometry parameters can therefore be determined, which are $\alpha_0=177.3_{-1.7}^{+1.3}$ degrees for the inclination angle and $\zeta_0=178.3_{-1.0}^{+0.8}$ degrees for the sight line angle, 
%. Therefore, $\beta=\zeta_0-\alpha_0\simeq 1(2)^\circ$ 
as listed in Table~\ref{table:geo}. This pulsar is an aligned rotator.

The sensitive FAST observation of PSR J1932+1059 also revealed the highly polarized continuous bridge emission between the interpulse and main pulse, with an intensity of only 1/1000 of the main pulse, see Figure~\ref{fig:Obs_profs}. Compared with previous results from the Arecibo observations at 1170~MHz \citep{mr04}, the trailing part of profiles in the phase range of $C_3$ has been clearly detected, and emission in the ranges of $E_0$ and $E_1$ can be seen. We use the piecewise RVM to fit the PA variations, and get the geometrical parameters of $\alpha_0=33.8_{-2.2}^{+1.6}$ and $\zeta_0=69.6_{-6.4}^{+5.5}$ degrees, 
%and hence $\beta=\zeta_0-\alpha_0\simeq 36(7)^\circ$,
demonstrating that this pulsar is an orthogonal rotator.

%==================================
\section{A model for polarized radio emission from a static pulsar magnetosphere}

To understand radio emission generated from any height at any rotation phase of pulsar magnetosphere, we here do simulations and examine the projected emission on the sky.

\subsection{Assumptions used in simulations}\label{sect:assum}

We assume that the magnetosphere of a pulsar is a static dipole, within which relativistic particles stream out from the two magnetic poles along the curved magnetic field lines and radiate via curvature
emission mechanism.

\subsubsection{Dipole geometry}
\label{sect:geo}

The magnetic field of a pulsar is assumed to be an inclined dipole field, 
\begin{equation}
\mathbf{B}=B_\star \left(\frac{R_\star}{r}\right)^3[3\hat{\mathbf{r}}(\hat{\mathbf{r}}\cdot \hat{\mathbf{m}})-\hat{\mathbf{m}}].
\end{equation}
Here, $B_\star$ and $R_\star$ are the surface magnetic field and
neutron star radius, $\hat{\mathbf{m}}$ is unit vector along the magnetic axis $\mathbf{m}$, $\hat{\mathbf{r}}$ represents the unit vector along $\mathbf{r}$ with an amplitude of 
\begin{equation}
  r=r_e \sin^2\theta.
\label{eq:r}
\end{equation}
Here $r_e$ is the field line constant and $\theta$ is the polar angle of a point on a given magnetic field line with respect to the magnetic
axis. Considering a magnetic dipole moment $\mathbf{m}$ inclined with an angle of $\alpha$ 
%
%\footnote{\add{In model calculation, we use the RVM convention for $\alpha$, $\beta$ and $\zeta=\alpha+\beta$. They relate to $\alpha_0$, $\beta_0$ and $\zeta_0=\alpha_0+\beta_0$ from observation convention by following $\alpha=180^\circ-\alpha_0$, $\beta=-\beta_0$ and $\zeta=180^\circ-\zeta_0$ \citep{ew01}.}}
%
with respect to the rotation axis $\mathbf{\Omega}$ as shown in Figure~\ref{fig:geo}, the parameterization of $\mathbf{r}$
and $\hat{\mathbf{m}}$ in a Cartesian coordinate system with the
z-axis along $\mathbf{\Omega}$ can be found from Equations (1),
(2) and (3) of \citet{gan04}. 

Pulsar radio emission is generally deemed to be generated from the
open magnetic field line region of pulsar magnetosphere. As bounded by
the light cylinder, the open field lines have a minimum of $r_e$
within a given magnetic field line plane $\phi$, which is denoted as
$r_{\rm e,lof}$ and can be calculated with the vector or projection methods. In the {\it Vector method}, the last open magnetic field lines are considered as a sphere that is tangential to the light cylinder. At the tangent point, the cross product of the differential vectors of $\mathbf{r}$ is perpendicular to the rotation axis, i.e., $(\frac{\partial 
  \mathbf{r}}{\partial \theta} \times \frac{ \partial
  \mathbf{r}}{\partial \phi})\cdot \hat{\Omega}=0$. By solving for the
polar angle $\theta_t$ of the tangent point, one can get $r_{\rm e,lof}$ as demonstrated by Equations (13), (14) and (15) of
\citet{gan04}.  In the {\it Projection method}, a curved magnetic field line is tangential to the light cylinder at the position $\mathbf{r_t}$, which has a component $r_p$ perpendicular to the rotation axis with a magnitude of $cP/2\pi$. The $r_{\rm e,lof}$ can be obtained by solving $d r_p/d \theta=0$, as depicted by Equations (7), (8) and (9) of \citet{zqh+07}.

\begin{figure}
  \centering
  \includegraphics[width = 0.45\textwidth] {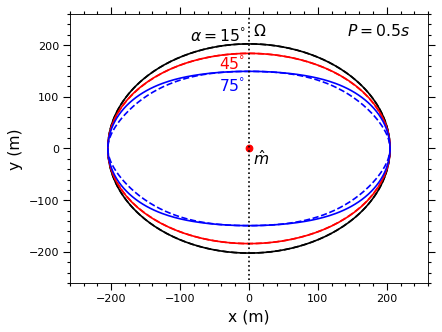}
  \caption{Polar cap shape for a pulsar rotating with a period of
    0.5s. The polar cap is centered around the magnetic axis
    $\hat{\mathbf{m}}$. The meridional plane defined by the magnetic
    and rotation axes is indicated by the black dotted line. The
    black, red and blue lines are for polar caps with
    $\alpha=15^\circ$, $45^\circ$ and $75^\circ$. The solid and dashed
    lines represent polar caps calculated with vector and projection
    methods, respectively.}
  \label{fig:capshape}
\end{figure}

Foot points of the last open magnetic field lines define the boundary of a polar cap on neutron star surface, which reads,
\begin{equation}
  (x_{\star,\rm lof}, y_{\star,\rm lof})= R_\star \theta_{\star,\rm lof} ( \cos \phi , \sin \phi).
\end{equation}
Here, $\theta_{\star,\rm lof} =\arcsin(\sqrt{R_\star/r_{\rm e,lof}})$, which varies
with the phases of magnetic field line planes. The shape of the polar cap is shown in Figure~\ref{fig:capshape} for different inclination angles for a pulsar with a rotating period of 0.5s. The polar cap has a size of about 200~m but is compressed in the meridinal plane due to the inclination of the magnetic axis from the rotation axes. % increases from $15^\circ$ to $75^\circ$. Both vector and projection methods give similar results for small inclination angles, but their discrepancy becomes large when $\alpha$ increases to $75^\circ$, as demonstrated in Figure~\ref{fig:capshape}. 
The polar cap calculated via the projection method is used in the following simulations.

\subsubsection{Distribution of relativistic particles}
\label{sect:den}

Relativistic particles are believed to be generated by sparking
processes at polar caps above the neutron star surface. It has been suggested that only some selected regions of the polar cap can spike rather than the entire cap. The preferred region might be organized in a central core and/or conal shape \citep{rs75}. The density and distribution of particles are much uncertain, although there are many simulations \citep[e.g.][]{cgc+21,pk22}. We assume that a bundle of particles is symmetrically distributed around the magnetic axis in a Gaussian shape,
\begin{equation}
  f_e(r,\theta,\phi)=\frac{f_0}{r^3}\exp\left[-\frac{(\theta_\star/\theta_{\star,\rm lof})^2}{2 \sigma_\theta^2}\right].
  \label{eq:fe}
\end{equation}
Here, $f_0$ is a normalization constant, $\theta_\star = \arcsin(\sin \theta \sqrt{R_\star/r} ) $ and represents the polar angle of a field line footed on the neutron star surface, $\theta_\star/\theta_{\star,\rm lof}$ ranges from
0 to 1 and here $\theta_{\star,\rm lof}$ varies with the azimuth angle $\phi$ (see Figure~\ref{fig:capshape}), $\sigma_\theta$ is the standard deviation for the distribution.
The particle density scales with $r$ by following $1/r^3$, similar as the Goldreich-Julian density \citep{gj69}. The particles are assumed to have a fixed Lorentz factor $\gamma$, since their energy loss caused by curvature radiation is negligible compared to the particle energy itself.

\subsubsection{Emission mechanism}
\label{sect:emi}

Relativistic electrons and positrons are injected into the magnetosphere from the polar cap. In the near magnetosphere, the magnetic fields are extremely strong and the life time for any transverse movement is negligibly small. Relativistic particles stream along the curved magnetic field lines and experience curvature radiation. In the far magnetosphere with weak magnetic field, relativistic particles would have pitch angles with respect to the magnetic field lines. The synchro-curvature radiation is the probable mechanism for the radiation \citep{zc95}.

Considering a relativistic particle bunch that has charge $q$ and locates at the position $\mathbf{r}$ %with the amplitude of Eq.(\ref{eq:r}) 
in a magnetosphere, with the instantaneous trajectory approximated by $\mathbf{r_c}$ in the circular path approximation, one can get the radiating electric field generated by these relativistic particles at frequency $\omega$ in direction $\mathbf{n}$ reads \citep{jac75},
\begin{equation}
  \mathbf{E}(\omega)=\frac{q e^{i \omega R_0/c}}{\sqrt{2 \pi} c R_0} \int_{-\infty}^{\infty} \frac{\mathbf{n}\times[(\mathbf{n}-\mathbf{v})\times \mathbf{a}]}{(1-\mathbf{n}\cdot\mathbf{v})^2} e^{i\omega(t-\mathbf{n}\cdot\mathbf{r}_{\rm c})/c} dt.
  \label{eq:E}
\end{equation}
Here, $\mathbf{v}$ and $\mathbf{a}$ represent the velocity in units of light speed $c$ and the acceleration of the relativistic particle bunch. $R_0$ is the distance from the circular path center to observer. For a movement along a field line, $\mathbf{r_{\rm c}}$, $\mathbf{v}$ and $\mathbf{a}$ can be found from Equations (11), (4) and (7) in \citet{wwh12}. In the current simulation, we set the rotation velocity to be zero, i.e, $v_r=0$, for the static magnetosphere.

%++++++++++++++++++++++++++
\begin{figure*}
  \centering
  \includegraphics[bb =  7 8 248 495, clip, height = 0.354\textheight] {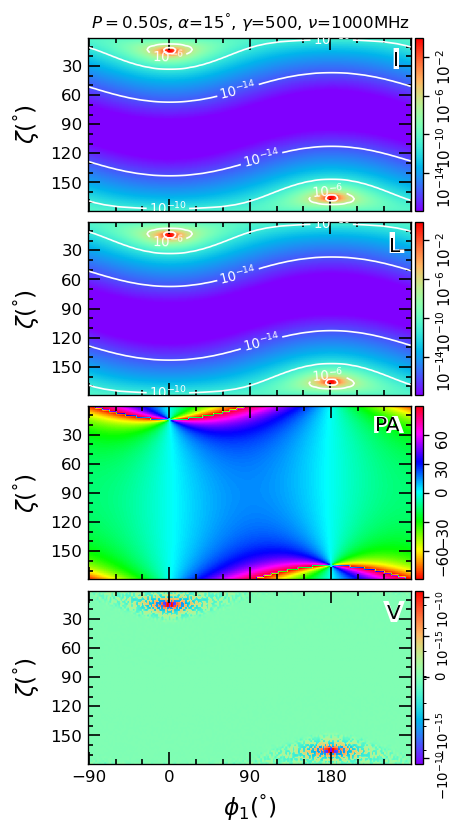}
  \includegraphics[bb = 51 8 248 495, clip, height = 0.354\textheight] {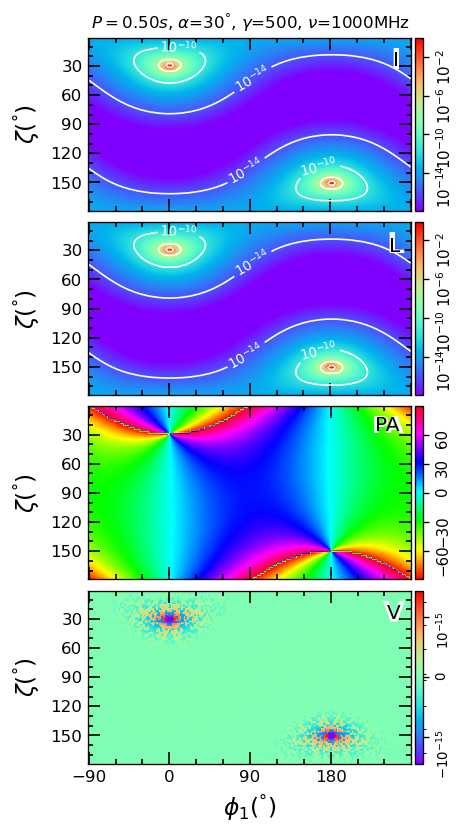}
  \includegraphics[bb = 51 8 248 495, clip, height = 0.354\textheight] {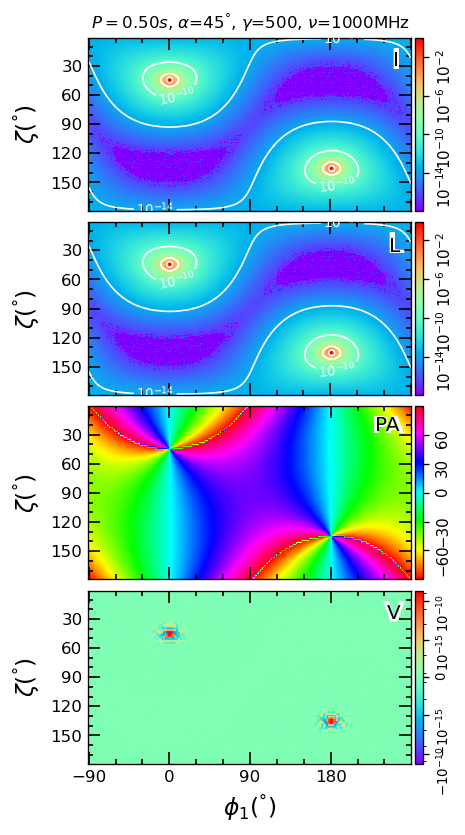}
  \includegraphics[bb = 51 8 248 495, clip, height = 0.354\textheight] {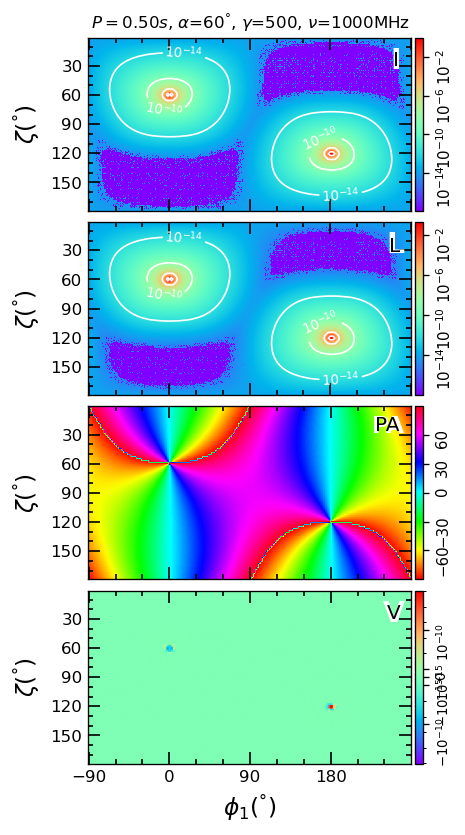}
  \includegraphics[bb = 51 8 275 495, clip, height = 0.354\textheight] {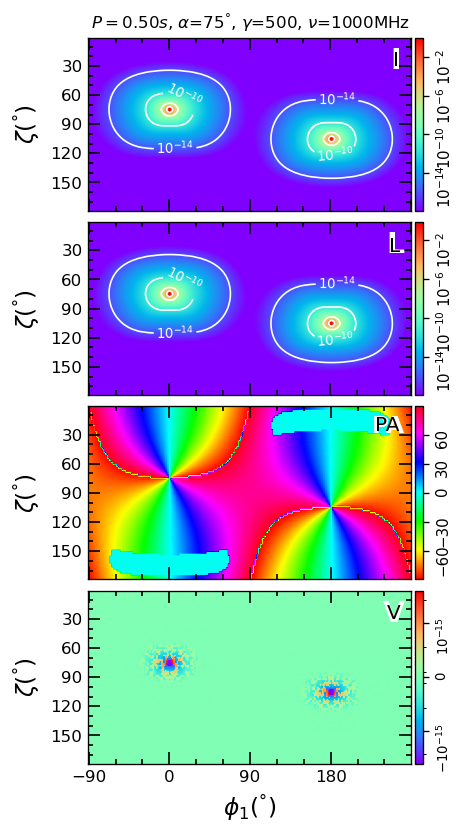}\\
  \caption{The Sky distribution of polarization emission intensity of a pulsar with different inclination angles $\alpha =15^\circ$, $30^\circ$, $45^\circ$, $60^\circ$ and $75^\circ$ (from left to right, as marked on the top). The distribution of the total intensity $I$, linear polarization $L$, polarization position angles $PA$, and circular polarization $V$ are presented from the top rank to the bottom. The other parameters used for simulations are $P=0.5$s, $\gamma=500$, $\sigma_\theta=0.2$, and $\nu=1000$ MHz.  }
  \label{fig:Patt_core_alpha_norot}
\end{figure*}

Pulsar emission should be coherent as implied by its high brightness temperature. The coherence might be caused by plasma waves that arrange electrons and positrons in phase like an antenna. When the arranged particles move along the curved magnetic field lines, the coherent curvature emission is generated \citep{cr77,ghw21}. Due to coherence, the emission intensity scales as $N^2E^2(\omega)$ instead of $N E^2(\omega)$, with $N$ representing the number of arranged particles. In our simulation, the coherence is treated by taking $q=f_e e$ in Eq.(\ref{eq:E}), and hence the emission intensity $\propto f^2_e$ in Eq.(\ref{eq:fe}). The emission intensity can be represented by the Stokes parameters via $I=E_x E_y^\star+ E_y E_y^\star$, $Q=E_x E_x^\star- E_y E_y^\star$, $U=2 Re[E_x^{*} E_y]$ and $V=2 Im[E_x^{*} E_y]$. Here, $E_x$ and $E_y$ are the two projected components of $E(\omega)$ in Eq.(\ref{eq:E}) on the x and y axes.

%------------------------
\subsection{Steps for simulations}\label{sect:steps}

We simulate radio emission from the pulsar magnetosphere and project it to the sky in the following steps.

{\it 1. Set up the pulsar magnetosphere:} For a pulsar with a given period $P$ and an inclination angle $\alpha$, the geometry and the distribution of relativistic particles in the whole magnetosphere are set up following the assumptions given in sections \ref{sect:geo} and \ref{sect:den}. At any location given by the coordinates $r$, $\theta$ and $\phi$, the density of relativistic particles $f_e$, their velocity $\mathbf{v}$ and acceleration $\mathbf{a}$ can be determined. 

{\it  2. Link the emission from a magnetic field tangent to the sky:} For a pulsar rotating at phase $\phi_1$, the emission generated at a height $r_0$ can be directed to a position in the sky. An observer in the sky defined by  the line of sight $\mathbf{n}$ can best receive the emission when $\mathbf{n}\cdot \hat{\mathbf{v}}=1$, due to relativistic beaming effect. The coordinate for the tangential emission point ($r_0$, $\theta_0$ and $\phi_0$) can be constrained by solving the equation.

{\it 3. Calculate emission for particles within a cone of $1/\gamma$:} The emission of a relativistic particle or a particle bunch is beamed primary in a cone with a size of $1/\gamma$ around $\mathbf{v}$. An observer in the sky at $\mathbf{n}$ can detect the emission from the cone. Therefore, the radiating electric field is calculated using Equation~\ref{eq:E} and the polarized emission is the incoherent sum of emission from the $1/\gamma$ cone of all particle bunches.

{\it  4. Get emission from the whole magnetosphere:} Radio emission at a given frequency can be generated from particles in a broad height region. We calculate the emission at a series of frequencies from all heights in pulsar magnetosphere.  

{\it  5. Project the emission onto the sky:} The emission at any frequency generated at any position inside the magnetosphere is projected onto the whole sky. An observer with a given line of sight receives the integrated emission from a series of $1/\gamma$ emission cones around all tangent points in the magnetosphere. The emission beams with larger intensities can be identified from the sky distribution of the summed emission.
 
{\it  6. Rotate the pulsar and get a profile for a given line of sight:} When a pulsar rotates, the projected emission in the sky rotates. In the observer's frame, the line of sight impacts the rotating sky in a given direction, and a pulse profile is obtained.

%++++++++++++++++++++++++++
\begin{figure}
  \centering
  \includegraphics[bb =  7 8 248 495, clip, height = 0.382\textheight] {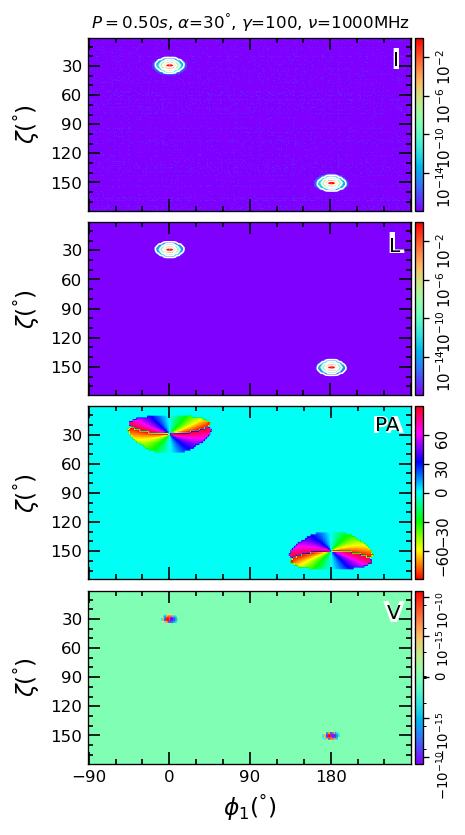}
  \includegraphics[bb = 51 8 275 495, clip, height = 0.382\textheight] {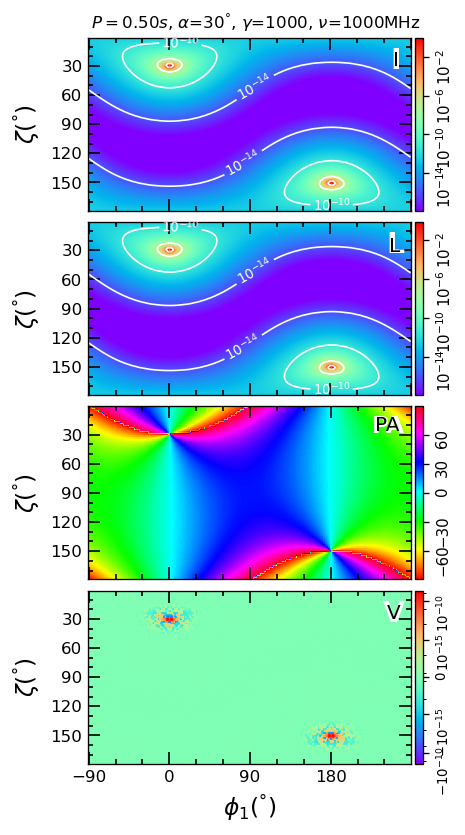}   \\
  \caption{The same as Figure~\ref{fig:Patt_core_alpha_norot} but for 
    relativistic particles with different Lorentz factors of $100$ and $1000$. The other parameters are fixed at $P=0.5s$, $\alpha=30^\circ$s and $\nu=1000$ 
    MHz.}
  \label{fig:Patt_core_alpha_norot_gamma}
\end{figure}
%--------------------------------

%++++++++++++++++++++++++++
\begin{figure}
  \centering
  \includegraphics[bb =  7 8 248 495, clip, height = 0.382\textheight] {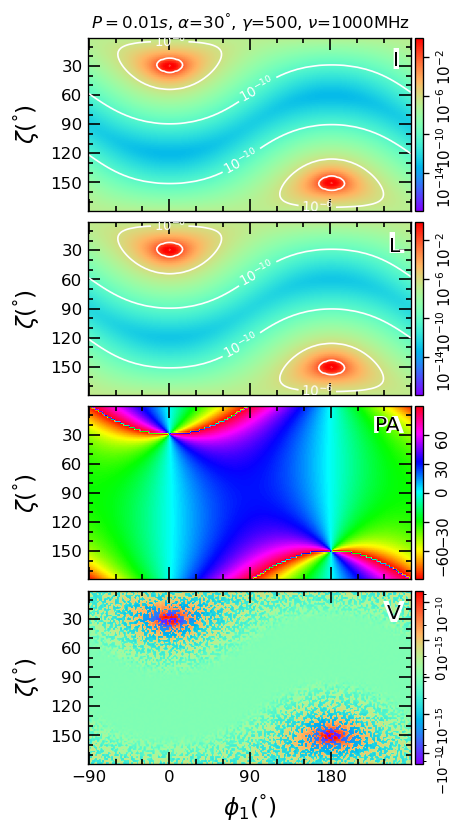}
  \includegraphics[bb = 51 8 275 495, clip, height = 0.382\textheight] {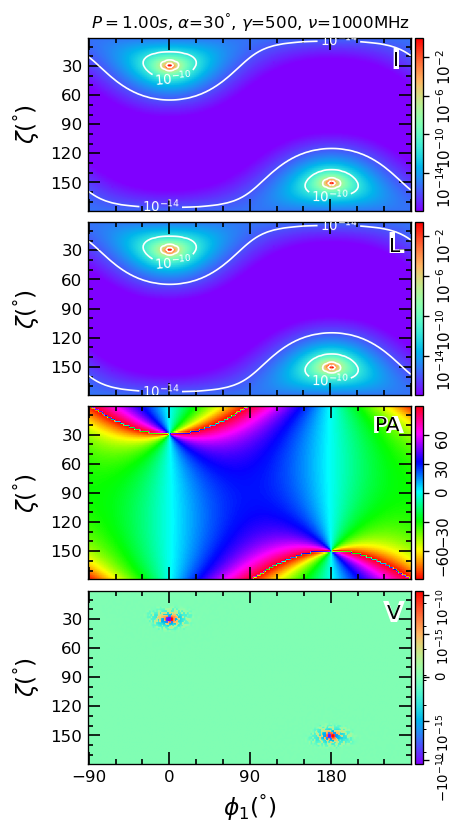}   \\
  \caption{The same as Figure~\ref{fig:Patt_core_alpha_norot} but for 
    different pulsar period of $0.01s$ and $1.0s$. The other parameters are fixed at $\alpha=30^\circ$s, 
    $\gamma=500$ and $\nu=1000$ MHz.  }
  \label{fig:Patt_core_alpha_norot_P}
\end{figure}
%--------------------------------

\subsection{Results}

%++++++++++++++++++++++++++
\begin{figure*}
  \centering
  \includegraphics[bb =  7 8 236 702, clip, height = 0.518\textheight] {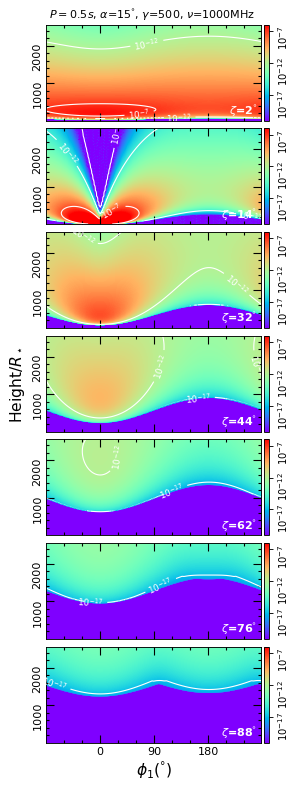}
  \includegraphics[bb = 42 8 236 702, clip, height = 0.518\textheight] {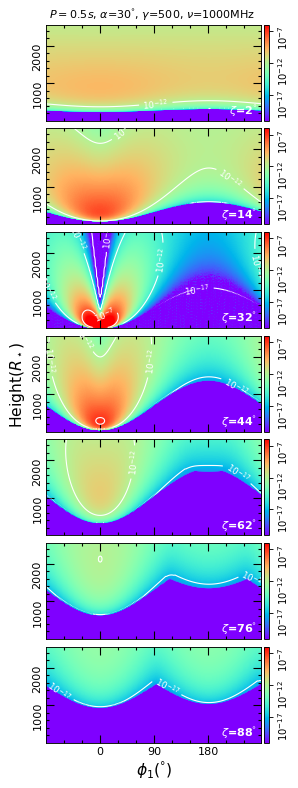}
  \includegraphics[bb = 42 8 236 702, clip, height = 0.518\textheight] {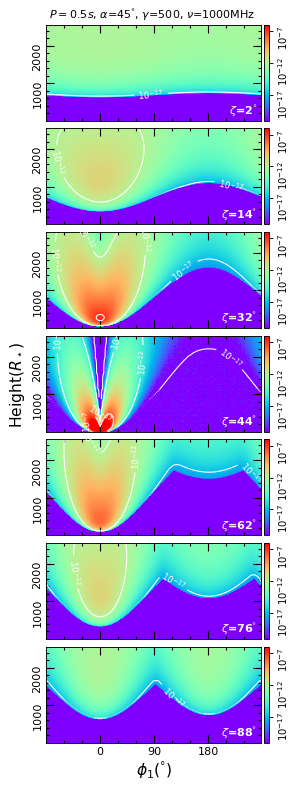}
  \includegraphics[bb = 42 8 236 702, clip, height = 0.518\textheight] {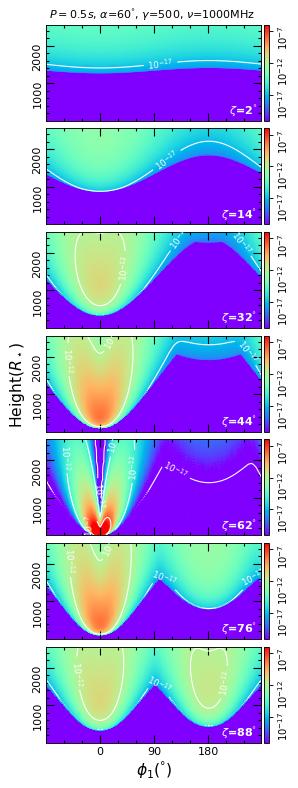}
  \includegraphics[bb = 42 8 260 702, clip, height = 0.518\textheight] {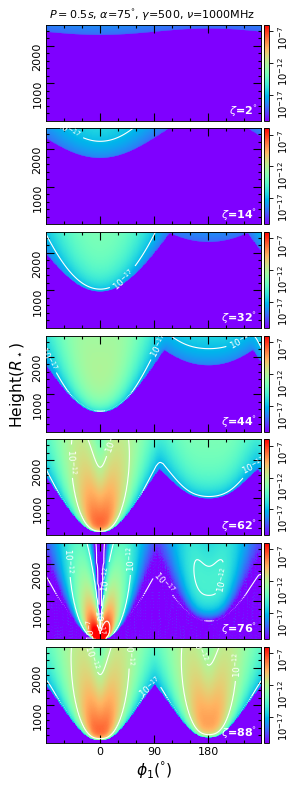}
  \caption{Simulated intensity distribution generated from different heights (the Y-axis) at different rotation phases (the X-axis) for different inclination angles of $\alpha = 15^\circ$, $30^\circ$, $45^\circ$, $60^\circ$ and $75^\circ$ (from the left to right) and different angles for the line of sight $\zeta=2^\circ$, $14^\circ$, $32^\circ$, $44^\circ$, $62^\circ$, $76^\circ$ and $88^\circ$ (from the top to bottom). Other parameters used for simulations are the same as those in Figures~\ref{fig:Patt_core_alpha_norot}. }
  \label{fig:Iheight}
\end{figure*}
%--------------------------------

\begin{figure*}
  \centering
  \includegraphics[bb =  9 9 282 1063, clip, height = 0.674\textheight] {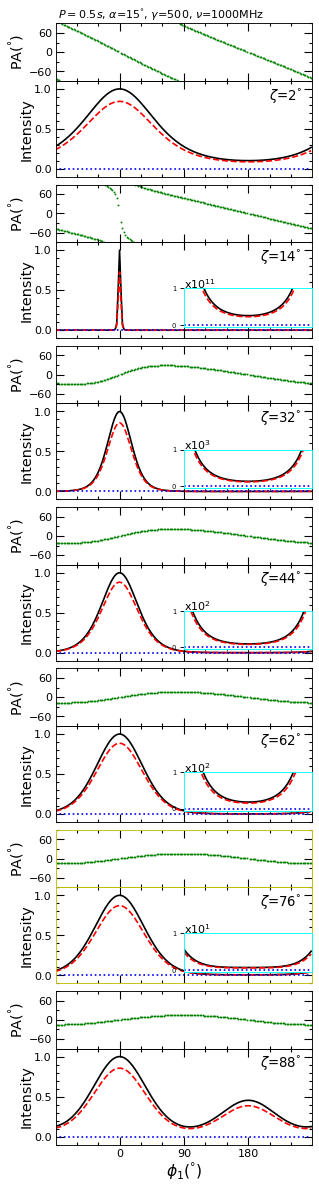}
  \includegraphics[bb = 51 9 282 1063, clip, height = 0.674\textheight] {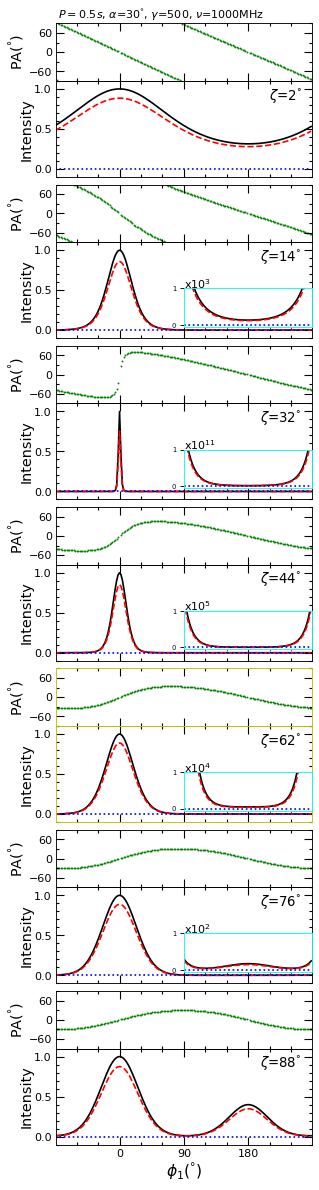}
  \includegraphics[bb = 50 9 282 1063, clip, height = 0.674\textheight] {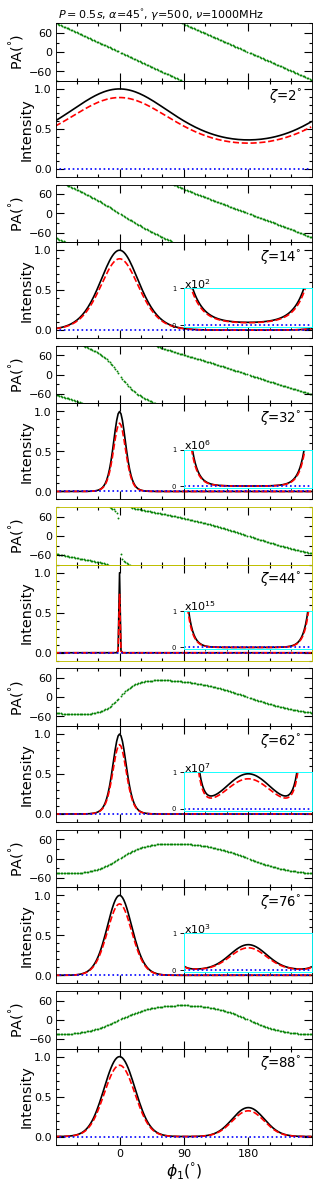}
  \includegraphics[bb = 51 9 282 1063, clip, height = 0.674\textheight] {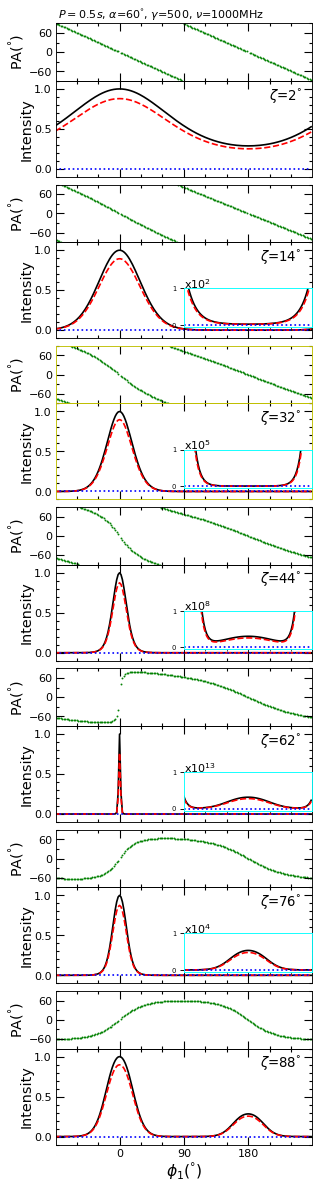}
  \includegraphics[bb = 51 9 283 1063, clip, height = 0.674\textheight] {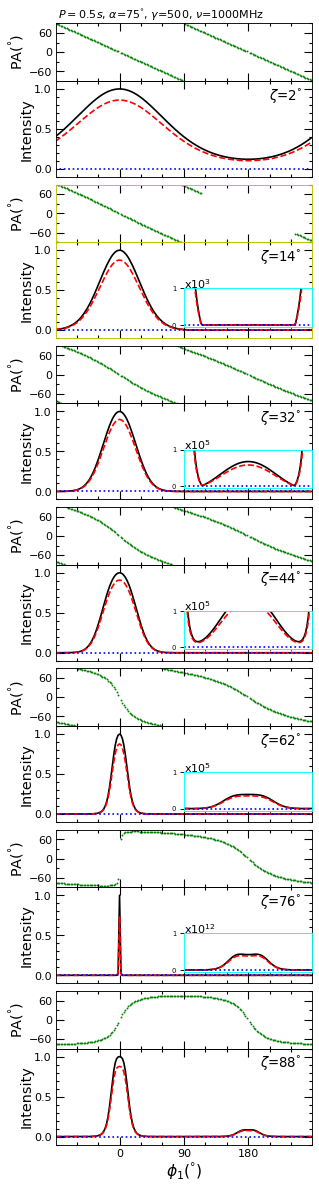}\\
  \caption{Pulsar polarization profiles from simulations with inclination angles of $\alpha= 15^\circ$, $30^\circ$, $45^\circ$, $60^\circ$ and $75^\circ$ from the left to right and line of sight at $\zeta=2^\circ$, $14^\circ$, $32^\circ$, $44^\circ$, $62^\circ$,  $76^\circ$ and $88^\circ$ from the top to the bottom. The total intensity $I$, linear polarization $L$, and circular polarization $V$ are represented by solid black lines, red dashed lines, and blue dotted lines in the lower sub-panels and the polarization position angles are given in the upper sub-panels. The inserted sub-panels show the enlarged part for $I$, $L$ and $V$ around the phase of $\phi_1=180^\circ$. These pulse profiles are calculated with the same sets of parameters as those in Figures~\ref{fig:Patt_core_alpha_norot} and \ref{fig:Iheight}. }
  \label{fig:Prof_core_alpha_norot}
\end{figure*}

By adopting the assumptions in Sect.~\ref{sect:assum} and following the steps in Sect.~\ref{sect:steps}, we obtain the sky maps for radio emission from the whole magnetosphere. They exhibit diverse emission patterns, as shown in Figures~\ref{fig:Patt_core_alpha_norot}, \ref{fig:Patt_core_alpha_norot_gamma}, and \ref{fig:Patt_core_alpha_norot_P}. 
When different lines of sight cut across these intensity distributions, various emission and pulse profiles can be detected, as shown in Figures~\ref{fig:Iheight} and \ref{fig:Prof_core_alpha_norot}. Indeed, one can have the radio emission across all rotation phases.

%++++++++++++++++++++++++++++++++++++++

\subsubsection{Patterns in the intensity distribution of emission generated from the whole magnetosphere}

Distributions of summed emission from the whole magnetosphere are shown in Figure~\ref{fig:Patt_core_alpha_norot} for pulsars with different inclination angles. No question that the strong emission is generated around two magnetic poles. The linear polarization is about $80\%$ the total intensity, and the %which distributes similarly as the total one. The 
position angles of linear polarization vary continuously with rotation phase. Because of the quasi-symmetric distribution of relativistic particles, the left and right-hand circular polarization are canceled seriously, so that the total circular polarization is only about $10^{-9}$ of the total intensity.

Radio emission generated from the whole magnetosphere emerges
at the entire rotation phases, despite with different inclination angles, though the intensity varies a lot. The emission at all phases is most preferably detectable by observers at a small angle from the rotation axis of a pulsar having a small inclination angle, i.e. viewing the almost aligned rotators near the pole. In such a case, the emission originates from one pole of the pulsar, e.g., $\zeta=2^\circ$ for a pulsar with $\alpha=15^\circ$. The emission at all phases might also be detected by an observer with the line of sight near $\zeta \simeq 90^\circ$, even though the emission is generated near the two poles of the quasi-orthogonal rotator, e.g. $\alpha=75^\circ$.

The patterns of emission intensity distribution are different for pulsars having their magnetosphere filled by relativistic particles of different energies $\gamma$, as shown in Figure~\ref{fig:Patt_core_alpha_norot_gamma}. The emission beams
surrounding both the magnetic poles are very small for a pulsar having less energetic particles, e.g. with Lorentz factor of 100, compared with the one having more energetic particles, e.g., with a Lorentz factor of 1000. This is because the more energetic particles remain to generate considerable curvature radiation from the higher magnetosphere where the curvatures of magnetic field lines are large. Moreover, polarization emission patterns are different for pulsars with different periods, as shown in 
Figure~\ref{fig:Patt_core_alpha_norot_P}. The emission of a pulsar with a short period of 0.01s, which resembles a millisecond pulsar, has a much wide beam compared to a long period pulsar of e.g. 1.0~s. This is because the magnetic field lines are largely bent for the pulsars with shorter periods.

%++++++++++++++++++++++++++
\begin{figure}
  \centering
  \includegraphics[width = 0.47\textwidth] {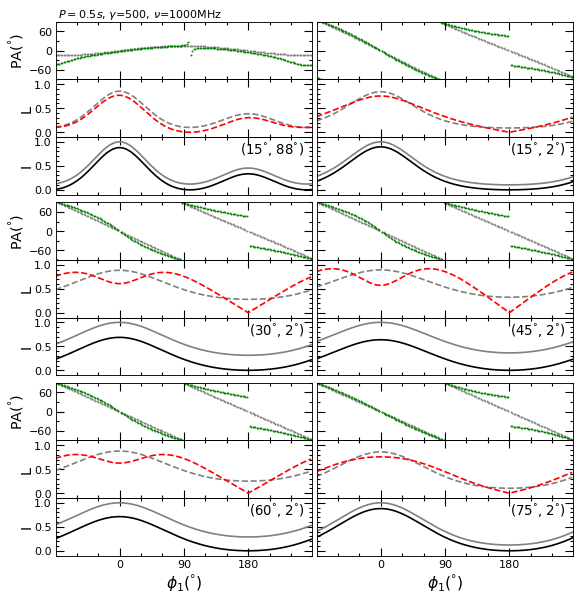}
  \caption{The same as Figure~\ref{fig:Prof_core_alpha_norot} but for polarized pulse profiles with persistent radio emission for pulsars with various geometries. If the persistent radio emission is not properly discounted, the $L$ and $PA$ curves would be distorted. The other parameters used for simulations are $P=0.5s$, $\alpha=30^\circ$s, and $\nu=1000$ MHz. }
  \label{fig:Prof_bias}
\end{figure}
%--------------------------

\subsubsection{Polarized pulse profiles}

When a given line of sight goes across the pulsar emission beam, the emission from a broad height range of pulsar magnetosphere can be detected. The emission intensity varies across heights, as shown in Figure~\ref{fig:Iheight}. In general, the line of sight with
small impact angles, i.e. a small $|\zeta-\alpha|$, can detect emission from the deep in the magnetosphere, as demonstrated by the emission detected by the line of sight of $\zeta=14^\circ$, $32^\circ$, $44^\circ$, $62^\circ$ and $76^\circ$ that cut pulsar emission beams with inclination angles of $\alpha =15^\circ$, $30^\circ$, $45^\circ$, $60^\circ$ and $75^\circ$. If the detectable emission comes from the other pole, in general, a weak profile component emerges at a rotating phase of $180^\circ$ from the main pulse, which is interpulse. When the impact angle gets
larger, emission from higher regions in the pulsar magnetosphere can be detected, and the intensity contrast between the main pulse and interpulse becomes smaller.

Because of the symmetry of emission from two poles of the pulsar magnetosphere, only emission above the equatorial plane is shown in Figure~\ref{fig:Prof_core_alpha_norot}. These polarized pulse profiles are obtained from the cutting of the emission patterns in Figure~\ref{fig:Patt_core_alpha_norot}, which is the sum of  emission from all possible heights as shown in Figure~\ref{fig:Iheight}. Radio emission from one 
magnetic pole is most likely detected for the pulsars with 
$(\alpha+\zeta) < 90^\circ$, as shown in the top-left sub-panels in
Figure~\ref{fig:Prof_core_alpha_norot}. For the pulsars with 
$(\alpha+\zeta) > 90^\circ$, radio emission from the two poles can be detected and the profiles are shown in the bottom-right sub-panels in Figure~\ref{fig:Prof_core_alpha_norot}. The weak emission in the entire rotation phase therefore could be detected from either an aligned or an orthogonal rotating neutron star, with the line of sight of $\zeta=2^\circ$ or $88^\circ$, regardless of the impact angles to the magnetic axis.

The gradients of position angle variations of polarized emission depend on the sign of impact angle of the line of sight. When the line of sight impacts the emission beam between the rotation and magnetic axes, i.e. the impact angle is negative, the polarization angles vary monotonically. When the line of sight impacts further than the magnetic axis with a positive $\beta$, as shown in Fig.~\ref{fig:geo}, the polarization angle variations have different senses for the main pulse and interpulse, as shown in the bottom-right sub-panels in Figure~\ref{fig:Prof_core_alpha_norot}. Vise versa for emission from the other hemisphere of pulsar magnetosphere.

Most striking is the persistent radio emission we obtained from the simulation, which is in addition to the pulsed radio emission and has not yet been considered in previous simulations. Generally, it is too weak to be detect for typical pulsar emission geometry. However, if the sight line is close to the rotation axes, e.g., $\zeta=2^\circ$, such a persistent emission can always be detected regardless of the inclination angle, for example $\alpha=15^\circ$, $30^\circ$, $45^\circ$, $60^\circ$ or $75^\circ$, as shown in Figures~\ref{fig:Prof_core_alpha_norot}. In addition, the persistent emission can be detected by sight lines with large angles, e.g. $88^\circ$ for a pulsar with a small inclination angle of $15^\circ$. In practice, such a persistent emission can only be measured by using the sensitive synthesis telescope for source detection after the pulsed emission is excluded. Note that such persistent emission is highly polarized. When it is ignored, the obtained pulse profiles would have a bias in polarization, as shown in Figure~\ref{fig:Prof_bias}.

%++++++++++++++++++++++++++
\begin{figure}[thbp]
  \centering
  \includegraphics[bb = 8 8 268 495, clip,  width = 0.34\textwidth] {J0953+0755_Simu-dis.pdf}
  \includegraphics[bb =  9 8 268 300, clip, width = 0.30\textwidth] {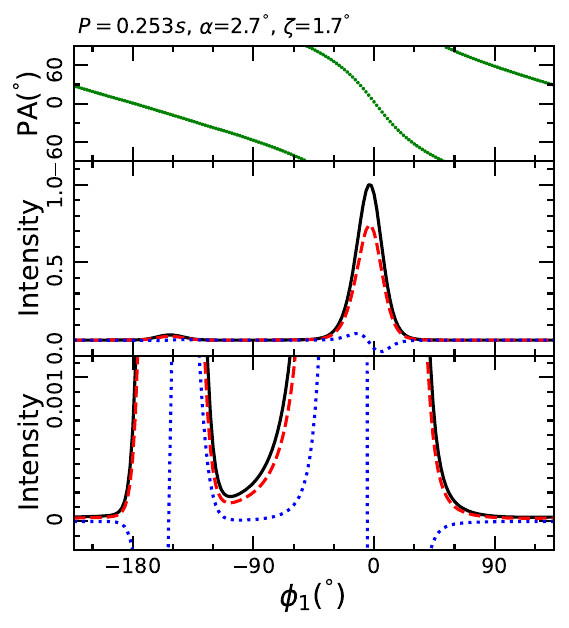}
   \caption{The specifically simulated sky distribution of polarization emission intensity for PSR J0953+0755 at $1250$ MHz. The simulation assumes two bundles of relativistic particles with $\gamma=500$. The sight line is at $\zeta=180^\circ-\zeta_0=1.7^\circ$. The obtained polarized pulse profile is given in the lower panels, similar to the observed one in Figure~\ref{fig:Obs_profs}. }
  \label{fig:comp0953}
\end{figure}

%++++++++++++++++++++++++++
\begin{figure}
  \centering
 \includegraphics[bb = 8 8 268 495, clip, width = 0.34\textwidth] {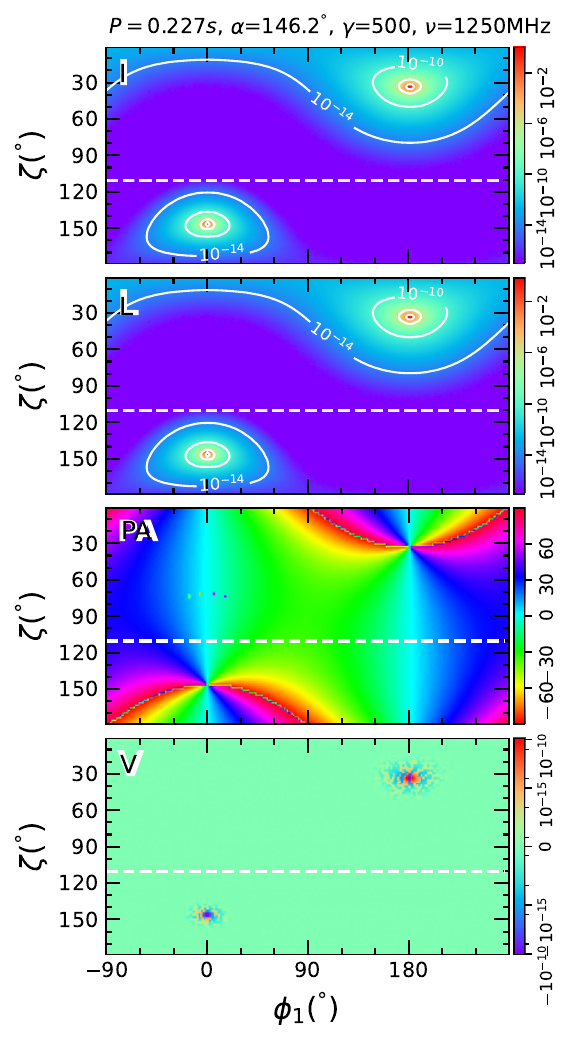}   \\
  \includegraphics[bb =  9 8 268 300, clip, width = 0.3\textwidth] {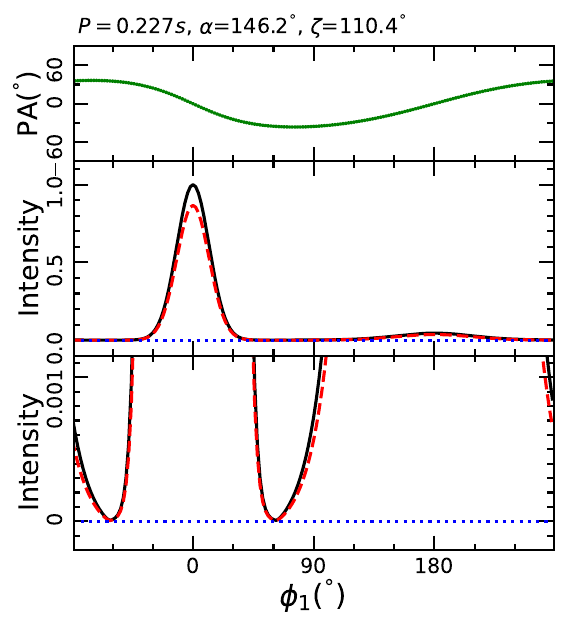}
  \caption{The same as Figure~\ref{fig:comp0953} but for PSR J1932+1059. The simulation assumes one bundle of relativistic particles with $\gamma=500$. The sight line is at $\zeta=180^\circ-\zeta_0 = 110.4^\circ$. }
  \label{fig:comp1932}
\end{figure}

\section{Comparing model with observations}

With the above simulations, one can see that the emission from all rotation phases can be generated. Here, we take simulations with specific parameters to generate the polarized radio emission mimic to the observed profiles of PSRs J0953+0755 from one pole of the magnetosphere and J1932+1059 from two poles .
%By assuming the energy and density distribution of relativistic particles, we model the sky distribution of polarized emission and pulse profiles. 

Simulations for PSR J1932+1059 take the relativistic particles from both poles with the distribution defined by Eq.~\ref{eq:fe} and  $\sigma_\theta=0.21$. The sky distribution of the emission generated by these particles and the polarized pulse profile are shown in Figures~\ref{fig:comp1932}. Simulations produced similar polarization  profiles and PA curve as the observed ones. Due to the symmetric distribution of particles around the magnetic axis, the circular polarization has the positive and negative senses canceled out. In fact, the FAST observations exhibit much more details for every individual pulses with intensity variations on short time scales and the diverse circular polarization and orthogonal modes. These features can not be reproduced by a simple model with a given set of parameters for a simple bundle of particles. All in all, our simulations reveal that emission can be generated in pulsar magnetosphere in all rotation phases. % Its improved understanding will be achieved by accounting for a much more dynamical magnetosphere and the propagation effects within which.  

%==================================
\section{Discussion and conclusions}
\label{sec:conclusions}

Sensitive radio observations show that pulsar emission emerges in the entire rotation phase. %Predictions of the model can be compared with observations by FAST. Emission of 
FAST observations of PSRs J0953+0755 and J1932+1059 have shown a huge dynamical ranges, and detected emission in some rotation phase down to less than 1/10000 the peak intensity. Their PA swings do not meet with the traditional RVM across the wide rotation phases. We developed a piecewise RVM which can be used to fit the PA values and get the emission geometrical parameters.

We have developed a theoretical model to understand the emission and polarization generated in the entire pulsar magnetosphere. We assumed the dipole geometry, symmetric distribution of relativistic particles, and the curvature emission mechanisms. The simulations have been carried out step by step, and the following conclusions can be derived:

1) Polarized radio emission can be generated in the entire rotation phases, though the intensity is much weak compared to the main emission beam around the magnetic poles. It is most likely produced by the pulsars having small periods and ones with the magnetosphere filled with much more relativistic particles. 

2) The emission can be detected from the entire phase for two kinds of geometry. One is for the sight lines close to the rotation axis. Where the pulsed emission, together with the persistent radio emission, are detectable regardless of the inclination angle of a pulsar. The other is for the sight lines close to the equator, where emission from an orthogonal rotating neutron star is most likely to be detected.  

3) Polarization angles of radio emission vary monotonically when the line of sight cuts the beam in between the rotation and magnetic axes. When the line of sight cuts the beam further than the magnetic axis, the senses of polarization position angle variation are opposite for the main pulse and interpulse. 

4) The persistent radio emission is highly polarized. The pulsar polarization profiles would be distorted if such emission is ignored. 

By assuming the energy and density distribution of relativistic particles and using the measured spin period and geometry parameters, we specifically simulate the polarized radio emission of PSRs J0953+0755 and J1932+1059. The model can matches observations for the main profile features.

Our model simulations only take the continuous injection of relativistic particles from the polar cap of the magnetic poles. Discrete injection due to sparking processes could lead to variable emission features at different time scales. Instead of static magnetosphere, many effects have not yet considered, such as asymmetric as caused by rotational sweepback effect \citep[e.g.][]{deu55, dh04}, or polar cap currents \citep{ha01}. In addition, the rotation itself has a great influence on the trajectory of particles and hence pulsar emission, as demonstrated by \citet[][]{bcw91, wwh12}. Even after the emission is generated, the propagation effects can also significantly affect the observed polarization states \citep[e.g.][]{ms77,wl07}. Moreover, different profile components might originate from different heights of pulsar magnetosphere, as indicated by the relative phase shift $d \phi_i$ from the piecewise RVM modelling. These influences should be considered together when the modeled polarization profiles is compared to the really observed ones from pulsars in future works.

\begin{acknowledgments}
%\acknowledgments
We thank the referee for careful reading and helpful suggestions.
P. F. Wang is supported by the National Natural Science
Foundation of China (No. 12133004), National Key R\&D Program of China
(No. 2021YFA1600401 and 2021YFA1600400) and the National SKA
program of China (No. 2020SKA0120200).
J. L. Han is supported by the National Natural Science
Foundation of China (No. 11988101).
\end{acknowledgments}

%\begin{appendices}

\restartappendixnumbering

% \captionsetup[figure]{name={\bf Figure~A}}
% \captionsetup[table]{name={\bf Table~A}}
\renewcommand\thefigure{A\arabic{figure}}
\renewcommand\thetable{A\arabic{table}}
\appendix

%++++++++++++++++++++++++++
\begin{figure}
  \centering
  \includegraphics[width = 0.85\textwidth] {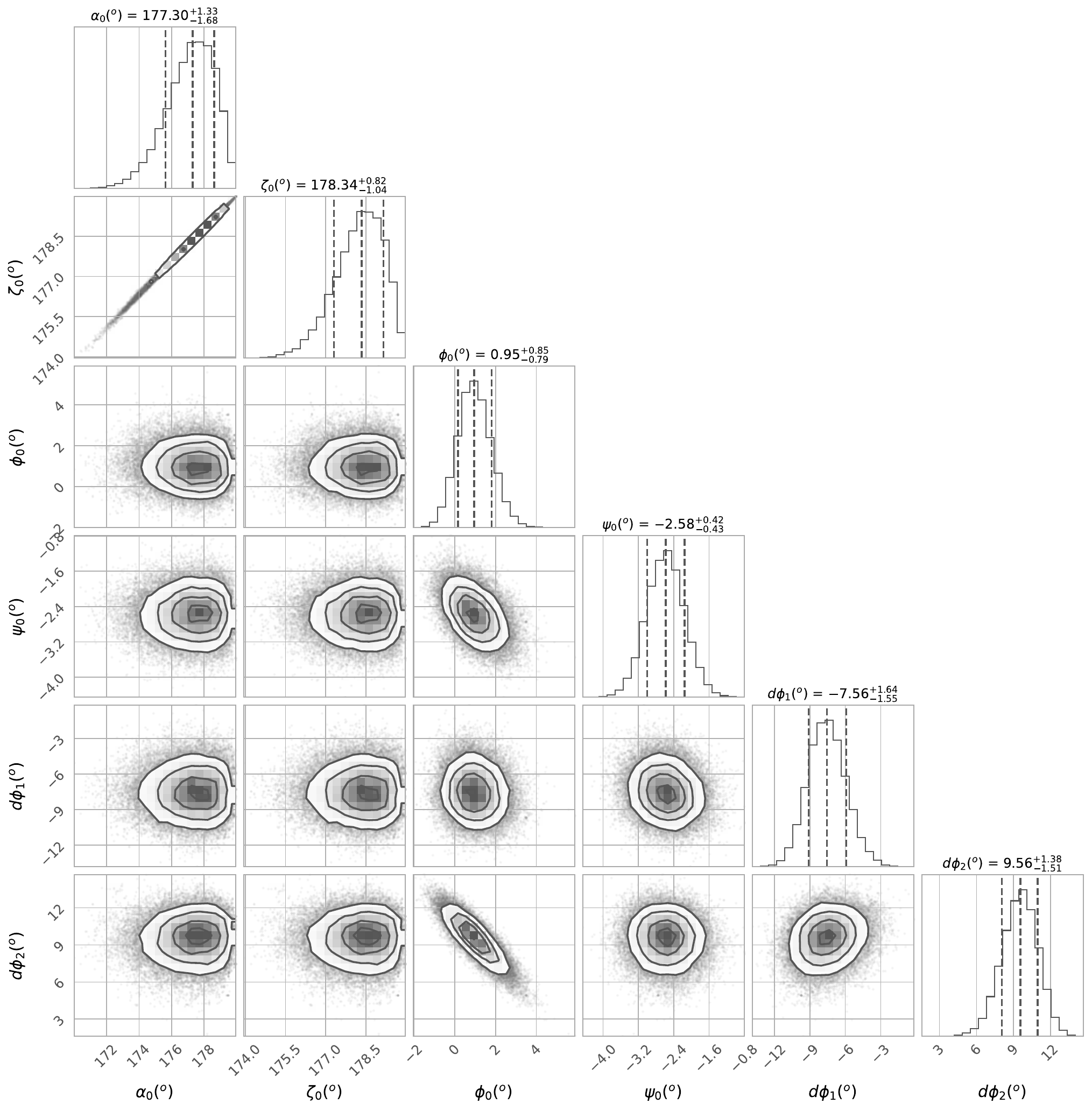}
  \caption{Marginalized probability distribution of the geometry parameters of PSR J0953+0755. The three dashed lines are for 16\%, 50\% and 84\% of the samples in the marginalized distributions for each parameter. From which, the parameter values and their uncertainties are estimated, as indicated at the top for each parameter.}
  \label{fig:mcmc_corner_J0953}
\end{figure}
\begin{figure}
  \centering
  \includegraphics[width = 0.85\textwidth] {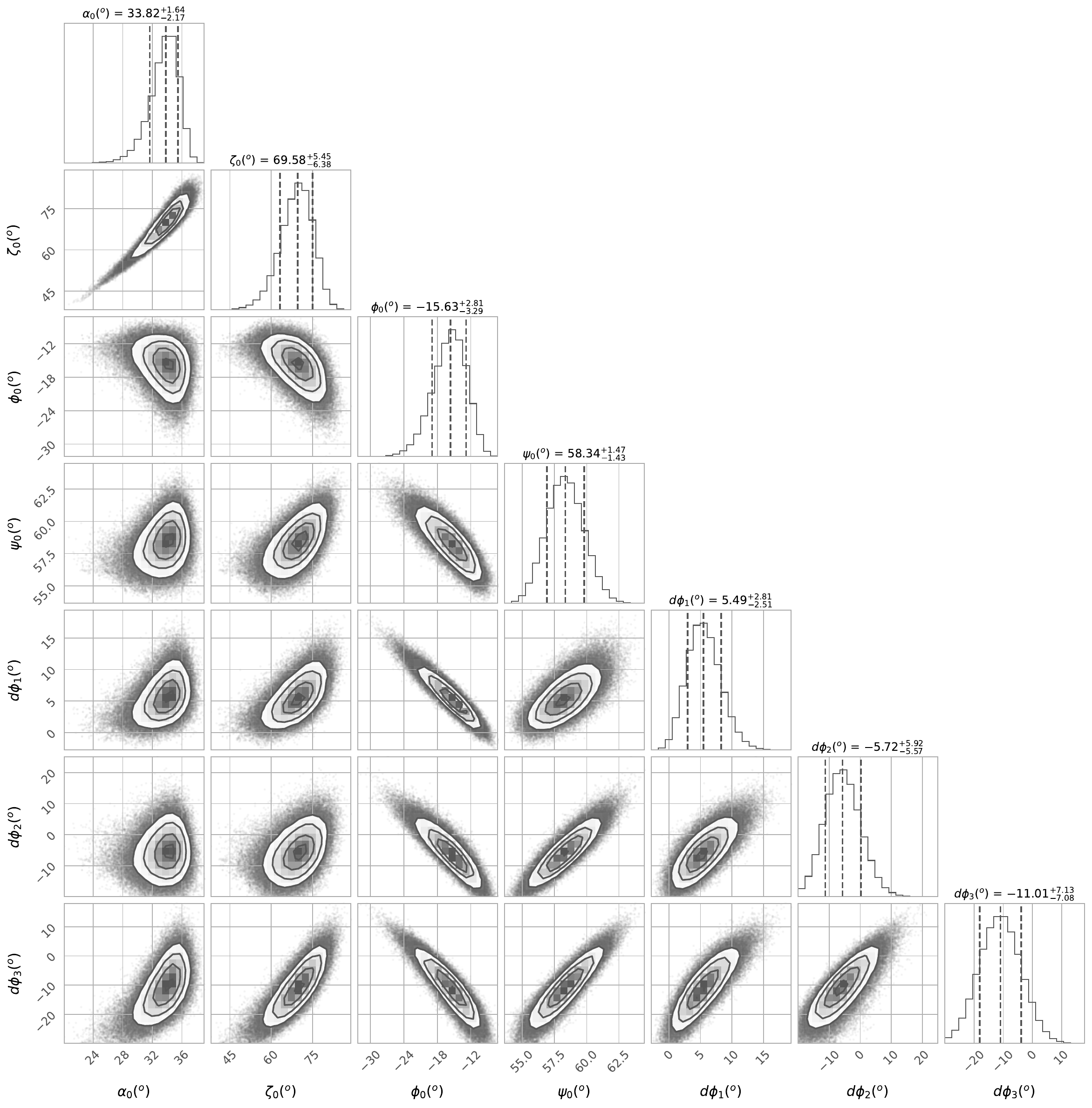}
  \caption{Same as Figure~\ref{fig:mcmc_corner_J0953} but for PSR J1932+1059 with four components modelled.}
  \label{fig:mcmc_corner_J1932}
\end{figure}

\section{The piecewise RVM model}
\label{sec:method}

For a given pulsar, a fixed set of geometry parameters, the inclination angle $\alpha$ and the impact angle $\beta$ or the sight line angle $\zeta$, can be determined by fitting the RVM to the observed position angle curve. However, for pulsars with wide profiles which have multiple widely spaced components generated from different heights in pulsar magnetosphere, one has to consider the aberration and retardation effects for different components, resulting in complex position angle swings. In fact, one can model the PA variation by using the following piecewise RVM function,
\begin{equation}
\psi=\left\{
\begin{array}{ccl}
\psi_0+\arctan[\frac{\sin \alpha \sin(\phi-\phi_0)}{\sin \zeta \cos\alpha -\cos\zeta\sin\alpha\cos(\phi-\phi_0)}] & &    {\rm  ~(for~main~pulse),} \\
\psi_0+\arctan[\frac{\sin \alpha \sin(\phi-\phi_0-d\phi_i)}{\sin \zeta \cos\alpha -\cos\zeta\sin\alpha\cos(\phi-\phi_0-d\phi_i)}]   & & {\rm  ~(for~i~th~component),} \\
\pi/2+\psi_0+\arctan[\frac{\sin \alpha \sin(\phi-\phi_0-d\phi_j)}{\sin \zeta \cos\alpha -\cos\zeta\sin\alpha\cos(\phi-\phi_0-d\phi_j)}]   & & {\rm  ~(for~j~th~component~with~orthogonal~mode).} \\
\end{array}
\right.
\label{eq:RVM_p}
\end{equation}
Here, $\alpha$, $\zeta$, $\phi_0$ and $\psi_0$ have the same meaning as the traditional RVM and can depict the PA of the main pulse component. $d\phi_i$ and $d\phi_j$ represent the relative phase shifts of the $i$th and $j$th profile components with respect to the main one in the model. If the PA of the $j$th component exhibits orthogonal mode jump with respect to the main pulse, a shift of $\pi/2$ is added to $\psi$.

We make use of the python packages, \textsc{emcee} and \textsc{corner}, and apply Equation~\ref{eq:RVM_p} to the observed data to get the estimated parameters. As shown in Figure~\ref{fig:mcmc_corner_J0953}, we model PA variations of the three phase ranges for PSR J0953+0755 ($C_0$, $C_1$ and $C_2$ marked in Figure~\ref{fig:Obs_profs}) and get the marginalized probability distributions of the parameters. Similar fitting has been made for  PSR J1932+1059 as well, as shown in Figure~\ref{fig:mcmc_corner_J1932} for the four components ($C_0$ to $C_3$ marked in Figure~\ref{fig:Obs_profs}). 

%\end{appendices}

\bibliography{psr_geometry}

\begin{thebibliography}{}
\expandafter\ifx\csname natexlab\endcsname\relax\def\natexlab#1{#1}\fi

\bibitem[{{Arons} \& {Barnard}(1986)}]{ab86}
{Arons}, J., \& {Barnard}, J.~J. 1986, \apj, 302, 120

\bibitem[{{Beskin} \& {Philippov}(2012)}]{bp12}
{Beskin}, V.~S., \& {Philippov}, A.~A. 2012, \mnras, 425, 814

\bibitem[{{Biggs}(1990)}]{big90}
{Biggs}, J.~D. 1990, \mnras, 245, 514

\bibitem[{{Bilous} {et~al.}(2022){Bilous}, {Grie{\ss}meier}, {Pennucci}, {Wu},
  {Bondonneau}, {Kondratiev}, {van Leeuwen}, {Maan}, {Connor}, {Oostrum},
  {Petroff}, {Verbiest}, {Vohl}, {McKee}, {Shaifullah}, {Theureau}, {Ulyanov},
  {Cecconi}, {Coolen}, {Corbel}, {Damstra}, {D{\'e}nes}, {Girard}, {Hut},
  {Ivashina}, {Konovalenko}, {Kutkin}, {Loose}, {Mulder}, {Ruiter}, {Smits},
  {Tokarsky}, {Vermaas}, {Zakharenko}, {Zarka}, \& {Ziemke}}]{bgp+22}
{Bilous}, A.~V., {Grie{\ss}meier}, J.~M., {Pennucci}, T., {et~al.} 2022, \aap,
  658, A143

\bibitem[{{Blaskiewicz} {et~al.}(1991){Blaskiewicz}, {Cordes}, \&
  {Wasserman}}]{bcw91}
{Blaskiewicz}, M., {Cordes}, J.~M., \& {Wasserman}, I. 1991, \apj, 370, 643

\bibitem[{{Cheng} \& {Ruderman}(1977)}]{cr77}
{Cheng}, A.~F., \& {Ruderman}, M.~A. 1977, \apj, 212, 800

\bibitem[{{Ching} {et~al.}(2022){Ching}, {Li}, {Heiles}, {Li}, {Qian}, {Yue},
  {Tang}, \& {Jiao}}]{clh+22}
{Ching}, T.~C., {Li}, D., {Heiles}, C., {et~al.} 2022, \nat, 601, 49

\bibitem[{{Cruz} {et~al.}(2021){Cruz}, {Grismayer}, {Chen}, {Spitkovsky}, \&
  {Silva}}]{cgc+21}
{Cruz}, F., {Grismayer}, T., {Chen}, A.~Y., {Spitkovsky}, A., \& {Silva}, L.~O.
  2021, \apjl, 919, L4

\bibitem[{{Desvignes} {et~al.}(2019){Desvignes}, {Kramer}, {Lee}, {van
  Leeuwen}, {Stairs}, {Jessner}, {Cognard}, {Kasian}, {Lyne}, \&
  {Stappers}}]{dkl+19}
{Desvignes}, G., {Kramer}, M., {Lee}, K., {et~al.} 2019, Science, 365, 1013

\bibitem[{{Deutsch}(1955)}]{deu55}
{Deutsch}, A.~J. 1955, Annales d'Astrophysique, 18, 1

\bibitem[{{Dyks}(2008)}]{dy08}
{Dyks}, J. 2008, \mnras, 391, 859

\bibitem[{{Dyks} \& {Harding}(2004)}]{dh04}
{Dyks}, J., \& {Harding}, A.~K. 2004, \apj, 614, 869

\bibitem[{{Everett} \& {Weisberg}(2001)}]{ew01}
{Everett}, J.~E., \& {Weisberg}, J.~M. 2001, \apj, 553, 341

\bibitem[{{Gangadhara}(2004)}]{gan04}
{Gangadhara}, R.~T. 2004, \apj, 609, 335

\bibitem[{{Gangadhara}(2010)}]{gan10}
---. 2010, \apj, 710, 29

\bibitem[{{Gangadhara} {et~al.}(2021){Gangadhara}, {Han}, \& {Wang}}]{ghw21}
{Gangadhara}, R.~T., {Han}, J.~L., \& {Wang}, P.~F. 2021, \apj, 911, 152

\bibitem[{{Gautam} {et~al.}(2022){Gautam}, {Ridolfi}, {Freire}, {Wharton},
  {Gupta}, {Ransom}, {Oswald}, {Kramer}, \& {DeCesar}}]{grf+22}
{Gautam}, T., {Ridolfi}, A., {Freire}, P.~C.~C., {et~al.} 2022, \aap, 664, A54

\bibitem[{{Gil} \& {Han}(1996)}]{gh96}
{Gil}, J.~A., \& {Han}, J.~L. 1996, \apj, 458, 265

\bibitem[{{Gil} \& {Snakowski}(1990)}]{gs90}
{Gil}, J.~A., \& {Snakowski}, J.~K. 1990, \aap, 234, 237

\bibitem[{{Goldreich} \& {Julian}(1969)}]{gj69}
{Goldreich}, P., \& {Julian}, W.~H. 1969, \apj, 157, 869

\bibitem[{{Gould} \& {Lyne}(1998)}]{gl98}
{Gould}, D.~M., \& {Lyne}, A.~G. 1998, \mnras, 301, 235

\bibitem[{{Han} {et~al.}(2009){Han}, {Demorest}, {van Straten}, \&
  {Lyne}}]{hdvl09}
{Han}, J.~L., {Demorest}, P.~B., {van Straten}, W., \& {Lyne}, A.~G. 2009,
  \apjs, 181, 557

\bibitem[{{Han} \& {Manchester}(2001)}]{hm01}
{Han}, J.~L., \& {Manchester}, R.~N. 2001, \mnras, 320, L35

\bibitem[{{Hankins} \& {Cordes}(1981)}]{hc81}
{Hankins}, T.~H., \& {Cordes}, J.~M. 1981, \apj, 249, 241

\bibitem[{{Hibschman} \& {Arons}(2001)}]{ha01}
{Hibschman}, J.~A., \& {Arons}, J. 2001, \apj, 546, 382

\bibitem[{{Jackson}(1975)}]{jac75}
{Jackson}, J.~D. 1975, {Classical electrodynamics}

\bibitem[{{Jiang} {et~al.}(2020){Jiang}, {Tang}, {Hou}, {Liu}, {Kr{\v{c}}o},
  {Qian}, {Sun}, {Ching}, {Liu}, {Duan}, {Yue}, {Gan}, {Yao}, {Li}, {Pan},
  {Yu}, {Liu}, {Li}, {Peng}, {Yan}, \& {FAST Collaboration}}]{jth+20}
{Jiang}, P., {Tang}, N.-Y., {Hou}, L.-G., {et~al.} 2020, RAA, 20, 064

\bibitem[{{Kou} {et~al.}(2021){Kou}, {Yan}, {Peng}, {Lu}, {Liu}, {Zhang},
  {Strom}, {Wang}, {Yuan}, {Yuen}, {Yu}, {Yao}, {Liu}, {Yan}, {Jiang}, {Jin},
  {Li}, {Qian}, {Yue}, {Zhu}, \& {FAST Collaboration}}]{kyp+21}
{Kou}, F.~F., {Yan}, W.~M., {Peng}, B., {et~al.} 2021, \apj, 909, 170

\bibitem[{{Large} {et~al.}(1968){Large}, {Vaughan}, \& {Wielebinski}}]{lvw68}
{Large}, M.~I., {Vaughan}, A.~E., \& {Wielebinski}, R. 1968, \nat, 220, 753

\bibitem[{{Lyne} \& {Manchester}(1988)}]{lm88}
{Lyne}, A.~G., \& {Manchester}, R.~N. 1988, \mnras, 234, 477

\bibitem[{{McLaughlin} \& {Rankin}(2004)}]{mr04}
{McLaughlin}, M.~A., \& {Rankin}, J.~M. 2004, \mnras, 351, 808

\bibitem[{{Melrose} \& {Stoneham}(1977)}]{ms77}
{Melrose}, D.~B., \& {Stoneham}, R.~J. 1977, Proceedings of the Astronomical
  Society of Australia, 3, 120

\bibitem[{{Nan}(2006)}]{nan06}
{Nan}, R. 2006, Science in China: Physics, Mechanics and Astronomy, 49, 129

\bibitem[{{Narayan} \& {Vivekanand}(1982)}]{nv82}
{Narayan}, R., \& {Vivekanand}, M. 1982, \aap, 113, L3

\bibitem[{{Narayan} \& {Vivekanand}(1983)}]{nv83}
---. 1983, \apj, 274, 771

\bibitem[{{Perry} \& {Lyne}(1985)}]{pl85}
{Perry}, T.~E., \& {Lyne}, A.~G. 1985, \mnras, 212, 489

\bibitem[{{Philippov} \& {Kramer}(2022)}]{pk22}
{Philippov}, A., \& {Kramer}, M. 2022, \araa, 60, 495

\bibitem[{{Phillips}(1990)}]{phi90}
{Phillips}, J.~A. 1990, \apjl, 361, L57

\bibitem[{{Pilkington} {et~al.}(1968){Pilkington}, {Hewish}, {Bell}, \&
  {Cole}}]{phb+68}
{Pilkington}, J.~D.~H., {Hewish}, A., {Bell}, S.~J., \& {Cole}, T.~W. 1968,
  \nat, 218, 126

\bibitem[{{Posselt} {et~al.}(2023){Posselt}, {Karastergiou}, {Johnston},
  {Parthasarathy}, {Oswald}, {Main}, {Basu}, {Keith}, {Song}, {Weltevrede},
  {Tiburzi}, {Bailes}, {Buchner}, {Geyer}, {Kramer}, {Spiewak}, \&
  {Krishnan}}]{pkj+23}
{Posselt}, B., {Karastergiou}, A., {Johnston}, S., {et~al.} 2023, \mnras, 520,
  4582

\bibitem[{{Qiao} {et~al.}(2004){Qiao}, {Lee}, {Wang}, {Xu}, \& {Han}}]{qlw+04}
{Qiao}, G.~J., {Lee}, K.~J., {Wang}, H.~G., {Xu}, R.~X., \& {Han}, J.~L. 2004,
  \apjl, 606, L49

\bibitem[{{Radhakrishnan} \& {Cooke}(1969)}]{rc69}
{Radhakrishnan}, V., \& {Cooke}, D.~J. 1969, \aplett, 3, 225

\bibitem[{{Rankin}(1983)}]{ran83}
{Rankin}, J.~M. 1983, \apj, 274, 333

\bibitem[{{Rankin}(1993)}]{ran93}
---. 1993, \apjs, 85, 145

\bibitem[{{Rankin} \& {Rathnasree}(1997)}]{rr97}
{Rankin}, J.~M., \& {Rathnasree}, N. 1997, Journal of Astrophysics and
  Astronomy, 18, 91

\bibitem[{{Ruderman} \& {Sutherland}(1975)}]{rs75}
{Ruderman}, M.~A., \& {Sutherland}, P.~G. 1975, \apj, 196, 51

\bibitem[{{Stairs} {et~al.}(1999){Stairs}, {Thorsett}, \& {Camilo}}]{stc99}
{Stairs}, I.~H., {Thorsett}, S.~E., \& {Camilo}, F. 1999, \apjs, 123, 627

\bibitem[{{von Hoensbroech} \& {Xilouris}(1997)}]{vx97}
{von Hoensbroech}, A., \& {Xilouris}, K.~M. 1997, \aap, 324, 981

\bibitem[{{Wang} \& {Lai}(2007)}]{wl07}
{Wang}, C., \& {Lai}, D. 2007, \mnras, 377, 1095

\bibitem[{{Wang} {et~al.}(2010){Wang}, {Lai}, \& {Han}}]{wlh10}
{Wang}, C., {Lai}, D., \& {Han}, J. 2010, \mnras, 403, 569

\bibitem[{{Wang} {et~al.}(2014{\natexlab{a}}){Wang}, {Pi}, {Zheng}, {Deng},
  {Wen}, {Ye}, {Guan}, {Liu}, \& {Xu}}]{wpz+14}
{Wang}, H.~G., {Pi}, F.~P., {Zheng}, X.~P., {et~al.} 2014{\natexlab{a}}, \apj,
  789, 73

\bibitem[{{Wang} {et~al.}(2012){Wang}, {Wang}, \& {Han}}]{wwh12}
{Wang}, P.~F., {Wang}, C., \& {Han}, J.~L. 2012, \mnras, 423, 2464

\bibitem[{{Wang} {et~al.}(2014{\natexlab{b}}){Wang}, {Wang}, \& {Han}}]{wwh14}
---. 2014{\natexlab{b}}, \mnras, 441, 1943

\bibitem[{{Wang} {et~al.}(2023){Wang}, {Han}, {Xu}, {Wang}, {Yan}, {Jing},
  {Su}, {Zhou}, \& {Wang}}]{whx+23}
{Wang}, P.~F., {Han}, J.~L., {Xu}, J., {et~al.} 2023, Research in Astronomy and
  Astrophysics, 23, 104002

\bibitem[{{Wang} {et~al.}(2022){Wang}, {Lu}, {Jiang}, {Lin}, {Lee}, {Liang}, \&
  {Xu}}]{wlj+22}
{Wang}, Z., {Lu}, J., {Jiang}, J., {et~al.} 2022, \mnras, 517, 5560

\bibitem[{{Xu} {et~al.}(2000){Xu}, {Liu}, {Han}, \& {Qiao}}]{xlhq00}
{Xu}, R.~X., {Liu}, J.~F., {Han}, J.~L., \& {Qiao}, G.~J. 2000, \apj, 535, 354

\bibitem[{{Zhang} {et~al.}(2007){Zhang}, {Qiao}, {Han}, {Lee}, \&
  {Wang}}]{zqh+07}
{Zhang}, H., {Qiao}, G.~J., {Han}, J.~L., {Lee}, K.~J., \& {Wang}, H.~G. 2007,
  \aap, 465, 525

\bibitem[{{Zhang} \& {Cheng}(1995)}]{zc95}
{Zhang}, J.~L., \& {Cheng}, K.~S. 1995, Physics Letters A, 208, 47

\end{thebibliography}

\end{document}